\documentclass[conference]{IEEEtran}

\usepackage{threeparttable}
\usepackage{graphicx}
\usepackage{url}
\usepackage{booktabs}
\usepackage{array}
\usepackage{amsmath}
\usepackage{xspace}
\usepackage{hyperref}
\usepackage{multirow}

\title{\bench: \\ A New Multi-Source Benchmark for Generalizable Astronomical Streak Detection}

\author{
\IEEEauthorblockN{Jiayou He}
\IEEEauthorblockA{
\textit{Interlake High School}\\
Bellevue, WA, USA\\
jiayou.he@outlook.com
}
\and
\IEEEauthorblockN{Jessica Yao}
\IEEEauthorblockA{
\textit{Interlake High School}\\
Bellevue, WA, USA\\
yaoxiaomei12@gmail.com
}
}

\newif\iffull

 \fullfalse   % Exclude full content

\newcommand{\bench}{\textsc{NEO-Bench}\xspace}

\usepackage{xcolor}
\newcommand{\yeye}[1]{\textcolor{blue}{#1}}
\newcommand{\jy}[1]{\textcolor{green!50!black}{#1}}

\begin{document}
\maketitle

\begin{abstract} Near-Earth Objects (NEOs) pose a threat that scales
with their size, ranging from regional damage to a global catastrophe, making the timely detection of NEOs crucial.
Moving NEOs can appear as faint streaks in long-exposure astronomical
images collected worldwide by thousands of telescopes at research and
citizen observatories. Detecting NEOs across these diverse sites requires
generalizable algorithms that can work reliably on astronomical images of 
varying quality, orientations, sky backgrounds, and noise levels. Although
algorithms have been developed to detect NEO streaks in images, existing 
detectors are commonly evaluated on data collected from a single source, making 
it difficult to determine whether they can generalize well and be reliably 
deployed at scale to thousands of sites without much additional manual tuning. 

To study the generalizability of existing streak-detection algorithms, we compiled 
a new benchmark, \bench, using 8,376 images collected from five different sources. The 
sources include the Hubble Space Telescope, a Stellina smart telescope, the United Arab 
Emirates Meteor Monitoring Network, a TETRA1 telescope using
a Celestron C14 with Fastar, and the Roboflow Asteroid dataset. We converted the
data to a common YOLO format and performed a manual audit to verify the quality
of the benchmark. We then defined within-source (in-domain) and leave-one-out 
(out-of-domain) evaluation protocols and conducted evaluations using classical, 
statistical, and deep-learning streak-detection methods such as Hough, Radon,
Gaussian PSF, and YOLO26L. Our results showed that leave-one-source-out $\boldsymbol{F_1}$ decreased in 14 of 20 image-level method--source pairs and 13 of 20 $\boldsymbol{\mathrm{IoU}@0.50}$ method--source pairs. In medium and hard datasets, $\boldsymbol{F_1}$ dropped in 11 of 12 image-level method--source pairs and 9 of 12 $\boldsymbol{\mathrm{IoU}@0.50}$ method--source pairs, indicating that cross-source performance remains inconsistent and highlighting an important direction for future research in streak detection.

The code and data used in \bench are publicly available on GitHub at
\url{https://github.com/he-jiayou/NEOBench} and on Hugging Face at \url{https://huggingface.co/datasets/jiayou-he/NEO-Bench}.
\end{abstract}

\section{Introduction}
Near-Earth Objects (NEOs) are asteroids and comets whose orbits approach
Earth's neighborhood. They pose a threat that scales
with their size, ranging from a localized ground impact to a planet-scale 
existential threat. Timely detection of NEOs is essential for planetary
defense because early discovery provides the time needed to characterize
potential impact hazards and, when necessary, implement mitigation strategies
\cite{nasa2023strategy}. In recent years, there have been multiple near-misses 
and unexpected NEO impacts. For example, in 2013, the Chelyabinsk meteor airburst 
injured more than a thousand people after entering Earth's atmosphere without prior
warning from astronomers or space-detection programs \cite{jpl2013fireball}. Another 
example is asteroid 2020 VT4, which, despite being 16--36 feet across, went undetected 
until 15 hours \emph{after} its closest approach. The asteroid passed just 232 miles above 
Earth's surface, approximately the altitude of the ISS, demonstrating the need for better
NEO detection to safeguard Earth \cite{irizarry2020asteroid}. %\jy{could remove extra info on asteroid 2020 VT4}

\textbf{Why is streak detection difficult?}
Modern astronomical surveys continuously generate terabytes of images,
creating a need for reliable automated detection systems.
Fast-moving objects can form elongated traces that are visually
distinctive when bright and long but difficult to detect when faint, short,
partially interrupted, or buried in sensor noise. These difficult cases can be 
confused with background stars and
instrumental artifacts \cite{duev2019deepstreaks}. In addition, many unrelated
phenomena—including satellite trails, meteors, cosmic rays, diffraction
spikes, clouds, and detector defects—produce visually similar structures,
requiring detection algorithms to achieve both high sensitivity and strong
robustness across diverse observing conditions.

% Classical pipelines remain useful because background subtraction,
% thresholding, morphology, and line transforms are fast and interpretable, but
% their fixed geometric and noise assumptions can be brittle under domain shift. 
% Section~II reviews these methods and their model-based and learned
% counterparts in more detail.
%DeepStreaks demonstrated that convolutional neural networks can efficiently
%filter fast-moving candidates in Zwicky Transient Facility (ZTF) observations
%. Parisot and Jaziri studied streak detection for smart-telescope imagery \cite{parisot2025detecting}, while Hubble Asteroid Hunter combined citizen-science labels with machine learning to discover 1,701 asteroid trails in archival Hubble observations
%\cite{kruk2022hubble}. Although 
Although advances in classical and deep-learning methods
\cite{duev2019deepstreaks,kruk2022hubble,parisot2025detecting}
(reviewed in Section~\ref{sec:related}) have significantly improved automated streak detection,
existing methods are predominantly evaluated using data from single sources
(e.g., a single observatory), providing limited evidence that they will
generalize to different telescopes, observing conditions, or labeling protocols.

\begin{figure*}[t]
\vspace{-10mm}
    \centering
    \includegraphics[width=0.90\textwidth]{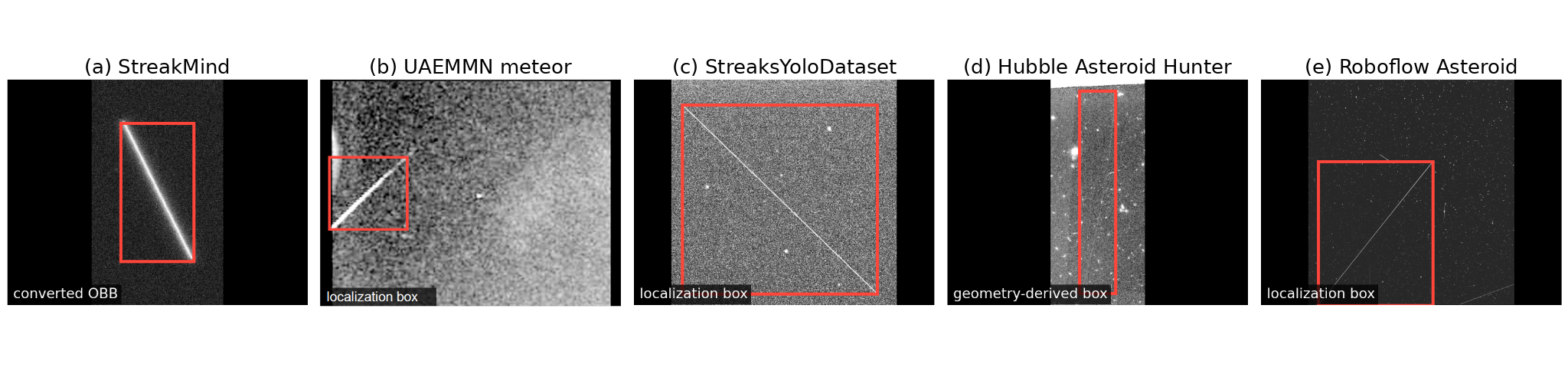}
    \vspace{-8mm}
    \caption{Representative examples from the five \bench sources:
    (a) StreakMind \cite{carrillo2026streakmind}, (b) UAEMMN \cite{alowais2023meteor}, (c) StreaksYoloDataset \cite{parisot2024dataset}, (d) Hubble Asteroid
    Hunter \cite{kruk2022hubble}, and (e) Roboflow Asteroid
    \cite{peek2024asteroid}. Red rectangles indicate ground-truth
    bounding boxes.
    Images are displayed with contrast
    stretching and cropping where necessary and are not shown at a common
    angular or pixel scale.}
    \label{fig:five_sources}
\end{figure*}

\textbf{The need for generalized models.} As of August 18, 2026, %\jy{do we need a date?} 
ESA’s NEO Coordination Centre listed 2,688 observatories that had contributed asteroid observations \cite{esaobservatories}. This extensive network encompasses a wide variety of telescopes, instruments, and observing conditions. As illustrated in Fig.~\ref{fig:five_sources}, images from different observatories can differ substantially in noise level, spatial scale, background density, and streak morphology, as well as in angular resolution, exposure time, point-spread function, preprocessing, compression, and sky brightness.

These variations create a significant \emph{domain shift}: training and test data may follow related but distinct distributions, meaning that strong in-domain performance does not necessarily transfer to previously unseen datasets~\cite{wang2022domain}. %Consequently, fairly comparing detection methods and assessing their ability to generalize across observatories remain challenging.

Nevertheless, cross-domain generalization is essential for practical streak detection. Training and maintaining a separate detector for every instrument and every observatory would be expensive, cumbersome, and unscalable. In contrast, a generalizable detector could be deployed across thousands of telescopes with minimal tuning, enabling more effective and scalable NEO searches and providing more comprehensive planetary defense. Existing evaluations of NEO detection methods do not adequately establish this capability.

\textbf{\bench}. To address this problem, we construct \bench, a unified benchmark for astronomical streak detection across multiple heterogeneous datasets.
\bench aggregates five publicly available sources into a common evaluation
framework, enabling systematic studies of cross-dataset generalization and
domain robustness. Rather than focusing on a single survey, the benchmark is
designed to measure how well detection methods transfer across different
instruments, imaging conditions, and labeling conventions.

\bench adopts a deliberately operational label definition:
a \emph{streak-positive} image contains an elongated trace accepted as
positive by its source dataset, regardless of whether the physical object is
an asteroid, satellite, debris object, cosmic ray, or meteor. These objects
are positive for the generic candidate-detection task because a high-recall
front end should forward them to later identification or rejection stages.
They are \emph{not} relabeled as confirmed NEOs. Since the current public
datasets do not provide consistent orbit-confirmed NEO-versus-non-NEO labels,
\bench evaluates the more fundamental task of \emph{astronomical streak
detection}. The project name reflects its planetary-defense motivation rather
than the specific labels being predicted.

% Representative examples from the five sources are shown in
% Fig.~\ref{fig:five_sources}. Together they span thin and broad streaks,
% different background densities and noise levels, and heterogeneous source
% labels, illustrating the visual diversity that motivates the benchmark.

Beyond assembling a benchmark, we systematically evaluated classical, statistical, and deep-learning detectors such as YOLO on \bench. The results revealed a separation between
within-source and leave-one-out results: performance generally decreased when methods were applied to an omitted
source, even with a larger, more diverse training set. \bench therefore identifies domain robustness, rather than only in-domain accuracy, as the
central challenge for future research.

\textbf{Contributions.} We make the following contributions:
\begin{enumerate}
    \item a multi-source meta-benchmark containing 8,376
    images from five sources with enhanced labels
    (e.g., bounding boxes added for UAEMMN);
    \item a task-aware unified representation that preserves source labels
    while supporting a common image-level streak-presence task;
    \item a cross-source evaluation protocol that measures transfer and
    generalization gaps and reveals the limitations of existing methods.
\end{enumerate}

\section{Related Work}
\label{sec:related}

\textbf{Existing Streak Detection Methods.}
Astronomical streak-detection methods broadly include classical line-search
methods, statistical image models, and deep-learning approaches. Classical
methods include the Hough and Radon transforms
\cite{duda1972hough,nir2018optimal}. Statistical methods fit explicit models
of streak-like sources, such as finite lines blurred by a point-spread
function \cite{veres2012trail}. Deep-learning methods can perform image
classification, object detection, or segmentation
\cite{redmon2016yolo,ronneberger2015unet,he2017maskrcnn}. While general-purpose foundation models such as GPT-5 can also perform streak detection, the sheer volume of astronomical data and GPT-5’s high per-call cost make large-scale deployment economically impractical -- a back-of-the-envelope calculation suggests that processing a single day’s worth of astronomical data with GPT-5 would cost many thousands of dollars.

%\yeye{May need to mention about GPT5 and other more general methods, say they can work effectively, but are too expensive (can quote an estimated cost number based on a rough calculation)}

\textbf{Existing NEO Benchmarks.}
To the best of our knowledge, \bench is the first publicly available benchmark that combines multiple 
astronomical-streak sources into one consistent bounding-box format.

%STar-DETR, https://pmc.ncbi.nlm.nih.gov/articles/PMC11859078/: "The images in this dataset
%are sourced from the Weihai Astronomical Observatory Xuanji Sky Survey Telescope and the 
%open access observational image data from the Observatorio Astronómico de La Sagra in Spain"

The five sources in \bench cover smart-telescope streaks (StreaksYoloDataset) \cite{parisot2024dataset}, Hubble
asteroid trails (Hubble Asteroid Hunter) \cite{kruk2022hubble}, labeled asteroid streaks (PEEK
Roboflow) \cite{peek2024asteroid}, satellite and NEO-like trails (StreakMind) \cite{carrillo2026streakmind}, and meteor/non-meteor
images (UAEMMN) \cite{alowais2023meteor}.
DeepStreaks was developed and assessed with data from the Zwicky Transient Facility (ZTF) at the Palomar Observatory in California \cite{duev2019deepstreaks,waszczak2017ptf}. 

% These efforts are complementary rather than directly comparable: their
% targets, sensors, positive definitions, label geometries, and negative
% examples differ. 
\bench is not intended as a sixth independent image source. Instead, it is a provenance-preserving meta-benchmark that integrates the five available sources while retaining their differences. It provides a shared schema for image-level streak presence and enables both within-source and cross-source evaluation. This design makes performance degradation on an unseen source measurable, something single-source benchmarks cannot reveal.

\section{Dataset Construction}

\subsection{Sources and Direct Inventory}

Table~\ref{tab:dataset_summary} reports the benchmark statistics. Images with streaks are labeled as ``streak-positive'' and have bounding-box labels, whereas images without streaks are labeled ``streak-negative.'' The
box total is the number of bounding boxes, not images. During conversion, we
standardized label and image names and converted existing labels to axis-aligned bounding boxes.

\begin{table*}[t]
\vspace{-2mm}
\centering
\caption{Inventory of the five \bench sources after dataset balancing. AABB denotes an
axis-aligned bounding box and OBB an oriented bounding box.}
\label{tab:dataset_summary}
\resizebox{\textwidth}{!}{%
\begin{tabular}{llrrrrrrrl}
\toprule
\textbf{Dataset} &
\textbf{Difficulty} &
\textbf{Train} &
\textbf{Val.} &
\textbf{Test} &
\textbf{Total} &
\textbf{Streak-positive} &
\textbf{Streak-negative} &
\textbf{Boxes} &
\textbf{Box/label processing} \\
\midrule
StreakMind
& Hard, background artifacts
& 1,484 & 278 & 94 & 1,856 & 840 & 1,016 & 858
& OBB polygons converted to AABBs \\
UAEMMN
& Hard, faint streaks
& 1,103 & 139 & 140 & 1,382 & 740 & 642 & 740
& Claude Opus 4.6 AABBs \\
StreaksYoloDataset
& Medium, bright streaks
& 1,722 & 333 & 333 & 2,388 & 1,642 & 746 & 1,642
& Original YOLO AABBs \\
Hubble Asteroid Hunter
& Easy, no negatives, artifacts
& 1,361 & 170 & 170 & 1,701 & 1,701 & 0 & 1,701
& WCS-projected endpoints; 5-pixel padding \\
Roboflow
& Easy, very few negatives
& 929 & 80 & 40 & 1,049 & 1,046 & 3 & 2,174
& Original YOLO AABBs \\
\midrule
\textbf{Combined}
& --
& \textbf{6,599} & \textbf{1,000} & \textbf{777}
& \textbf{8,376} & \textbf{5,969} & \textbf{2,407}
& \textbf{7,115} & -- \\
\bottomrule
\end{tabular}%
}
\vspace{-4mm}
\end{table*}
\textbf{StreakMind.}
The oriented bounding box (OBB) archive has 1,856 images, including 840 images with 858 streak
instances produced within a larger satellite-identification pipeline. For a
common baseline, each polygon is converted to the tight enclosing AABB. The
images came from La Sagra Observatory, Granada, Spain, using the TETRA1
telescope---a Celestron C14 with Fastar \cite{carrillo2026streakmind}. The streaks are relatively faint and hard to detect for classical methods. There are many background artifacts present in the images.

\textbf{UAEMMN.}
The original archive contains 7,400 meteor and 7,400 non-meteor images
associated with the United Arab Emirates Meteor Monitoring Network study
\cite{alowais2023meteor}. We sampled 10\% of UAEMMN to match the scale of the
other datasets and removed images that contained streaks but were labeled as
negative. This process resulted in 1,382 images: 1,103 for training, 139 for
validation, and 140 for testing. Because the original UAEMMN images did not
include bounding-box information, we used Claude Opus 4.6 to generate localized
streak bounding boxes for \bench and manually audited samples to check their quality. The images have some background artifacts, which can lead to false positives.

\textbf{StreaksYoloDataset.}
This source was created by Olivier Parisot using a Stellina smart telescope between March 2022 and February 2023 in the Greater Region of Luxembourg
 \cite{parisot2024dataset}. The local
archive has 2,388 images and 1,642 detector-ready bounding boxes. Its broad
streak class may include satellites, debris, and cosmic rays. Its streaks are bright, well-defined, and easy to detect. %Its split is approximately 72.1/13.9/13.9\% for train/validation/test.

\textbf{Roboflow Asteroid.}
The local reconstruction of the PEEK Roboflow Universe project contains 1,049
accessible version-5 records with YOLO-style asteroid-streak boxes. However,
the Roboflow page did not reveal where the data came from. The project credits
PEEK and specifies CC BY 4.0 \cite{peek2024asteroid}. The test split does not include negative images.

\textbf{Hubble Asteroid Hunter.}
Kruk et al.\ used citizen science and archival analysis to catalogue 1,701
asteroid trails in 1,316 unique Hubble observations \cite{kruk2022hubble}. Catalogued sky-coordinate
endpoints are projected into cutout pixels with the image world coordinate
system (WCS); the enclosing axis-aligned bounding box (AABB) then receives
five pixels of padding. This is reproducible geometry-derived supervision,
not a manually drawn rectangle. The dataset does not include negative images either.

\iffull 
    disproportionately influencing model training and aggregate results because of dataset size alone. Claude Opus 4.6 generated localized meteor bounding-box
    labels for this subset; these labels are a contribution of
    \bench rather than labels supplied by the original archive. Claude Opus 4.6
    examined each image, identified whether a meteor streak was present, and drew
    a localized axis-aligned bounding box around each identified streak. The box
    coordinates were normalized to the image dimensions and stored directly in
    the common YOLO representation. \yeye{shorten}
\fi

\subsection{Benchmark Tasks}

\bench defines:
\begin{itemize}
    \item \textbf{Image classification:} presence or absence of a streak-like
    target across all five sources using their available labels. It is a binary yes/no classification task.
    \item \textbf{Object detection:} detection of streak-like targets in
    the five sources with bounding boxes to assist human inspection and verification. % totaling 8,376 images and 7,115 boxes. %Both positive and negative images are retained. Detects the precise boundaries of streaks.
% \yeye{Right now the descriptions of these two tasks above look very similar, and hard for readers to tell their differences. It is better to move image classification first, say it is a binary yes/no classification task, then explain object-detection, and say that on top of yes/no classification object-detection needs to detect to precise boundary of streaks to assist human verification, so that readers can understand the differences.}
\end{itemize}

\iffull
    \subsection{Quality Control and Leakage Prevention}
    
    For every converted image, we recorded the original and converted file
    locations, the assigned split, and the conversion method in a manifest. We
    limited bounding-box values to the valid YOLO coordinate range and rejected
    labels that had no matching image. When a source did not provide an official
    split, we randomly assigned its images to the training, validation, and test
    sets using 42 as a fixed random starting value, ensuring that each run
    produces the same split.
    
    A purely random split can place closely related observations in both the
    training and test sets. For example, Hubble may contain several trails from
    the same observation, while ground-based sources may contain consecutive
    frames from one observing session. If these similar images appear in both
    sets, test performance may seem better than it would be on genuinely new data
    \cite{figueiredo2024leakage}. To prevent this leakage, we kept images from the
    same observation ID or capture session together in a single training,
    validation, or test split.
\fi

\subsection{Manual Audit and Label Verification}

In order to ensure data and label quality, each author independently reviewed 100 positive images randomly sampled
from each source, for a total of 500 images. Each image was inspected to
verify its label quality, with ambiguous cases that are potentially incorrect flagged in the process.

The first author flagged a total of 16 sampled images with positive labels as potentially negative, whereas
the second author classified 8 as negative. All 8 images identified as
negative by the second author were also identified by the first. These cases were visually ambiguous, which likely contributed to the
disagreement. The authors therefore agreed on 492 of the 500 images, corresponding to
98.4\% observed agreement. The Matthews correlation coefficient was 0.70,
and Gwet's AC1 was 0.98.

%Ambiguous cases were flagged, but no formal adjudication was performed. 
Overall, the percentages of sampled images flagged
were 1\% for StreaksYoloDataset, 3\% for Hubble Asteroid Hunter, 2\% for
Roboflow Asteroid, 0\% for UAEMMN, and 6\% for StreakMind, showing strong label quality of \bench despite the presence of a small fraction of ambiguous images.  

% \yeye{add details of the review process. Report aggreement between two labelers, e.g. refer to \url{https://www.innovatiana.com/en/post/inter-annotator-agreement} and report using metrics mentioned there}

\section{Experimental Evaluation}

The study addressed two research questions:
\begin{enumerate}
    \item \textbf{RQ1:} Which evaluated method performs best for within-source astronomical streak detection?
    \item \textbf{RQ2:} How well do models generalize to unseen sources?
\end{enumerate}

\subsection{Experimental Settings}

\textbf{Within-source} experiments trained or calibrated the methods on one source and evaluated them on that source's test split.

%\yeye{better to move ``V: Methods compared'' above ``experimental settings'' (as section IV.A), otherwise when you say YOLO26L and its hyperparameters below, people will find it abrupt. First introduce Yolo26L, then talk about hyper-parameters. Hyper-parameters can probably go inside YOLO26L section of ``Methods compared'', instead of being talked about in ``Leave-one-out'' here.} 

\textbf{Leave-one-out} experiments trained or calibrated the
methods on all but one source and evaluated them on the held-out dataset. 

\subsection{Metrics}
We report precision, recall, and $F_1$ for each dataset. For image-level evaluation,
an image is classified as positive when at least one candidate passes the
confidence threshold. Precision is the proportion of predicted-positive images
that are streak-positive, recall is the proportion of streak-positive images
that are detected, and $F_1$ is the harmonic mean of precision and recall
\cite{carrillo2026streakmind}. %\jy{we could add the math here to make it clearer}
For model $m$ and source $d$, define the signed $F_1$ change as
\begin{equation}
\Delta F_{1,m,d}
=
F_{1,m,d}^{\mathrm{LOO}}
-
F_{1,m,d}^{\mathrm{within}} ,
\end{equation}
so a negative value indicates worse performance after withholding the test
source. Reporting the full
set of within-source and leave-one-out results makes the effect of withholding
each source visible.

To integrate localization into our results, we used $\mathrm{IoU}@0.50$. 
Intersection over Union ($\mathrm{IoU}$) measures the overlap between a
predicted box $B_p$ and a ground-truth box $B_g$:
\begin{equation}
\mathrm{IoU}(B_p,B_g)
=
\frac{\operatorname{area}(B_p\cap B_g)}
{\operatorname{area}(B_p\cup B_g)}.
\end{equation} 
$\mathrm{IoU}@0.50$ marks a detection as a true positive only when the predicted box overlaps the ground-truth box by at least 50\%.
To get $\mathrm{IoU}@0.50$ results from the classical and statistical methods, Hough detected finite line segments directly, while Radon and Gaussian PSF first detected possible lines and then used image brightness to estimate their endpoints. Each line was widened and converted into a rectangular box. %to be continued, unless detector settings clarified earlier
\section{Methods Compared}

We evaluated one statistical and two classical methods spanning line voting,
projection-based matched filtering, and explicit image-formation modeling,
together with YOLO26L as a learned object detector.

\textbf{Hough.} The Hough transform applies signed, multi-scale directional filters and votes aligned responses in line-parameter space \cite{duda1972hough}. Candidates are scored using center-line and sideband responses, followed by endpoint recovery and coherence testing.

\textbf{Radon.} The Radon transform pads each image and integrates signed responses along trial lines \cite{nir2018optimal}. It searches a coarse angle grid, locally refines promising angles, and applies the same endpoint and coherence tests as Hough.

\textbf{Gaussian PSF.} The Gaussian PSF method fits a finite Gaussian-broadened line to Hough and Radon proposals, refining position, angle, length, width, amplitude, and background \cite{veres2012trail}. Detections must satisfy both likelihood-gain and line-coherence criteria.

\textbf{YOLO26L.} This deep-learning method directly predicts streak
locations and confidence scores from labeled images
\cite{jocher2026yolo26}. Within-source models were trained separately on each source's training split,
with its validation split used for model selection. Leave-one-source-out models
were trained on the training splits of all non-held-out sources, with their
validation splits used for model selection. The held-out source was used only
for testing.

YOLO26L was initialized from the pretrained \texttt{yolo26l.pt} checkpoint and fine-tuned independently for 30 epochs using a batch size of 16, an input size of $640\times640$, and a random seed of 0. Ultralytics resized each image to fit within the $640\times640$ canvas while preserving its aspect ratio, then padded the remaining pixels. %(letterboxing)
Each YOLO26L run started from the same pretrained \texttt{yolo26l.pt} checkpoint rather than from a previous \bench run. Using Ultralytics 8.4.3, automatic optimizer selection (\texttt{optimizer=auto}) chose MuSGD with an effective initial learning rate of approximately 0.002, momentum of 0.90, and weight decay of 0.0005. Training was performed on an NVIDIA A100 80~GB PCIe GPU. YOLO26L training and augmentation configurations are available in the public GitHub repository.

Test evaluation used \texttt{best.pt}, selected using Ultralytics' detection fitness based on validation $\mathrm{mAP}_{50:95}$. For image-level evaluation, $F_1$-optimal validation thresholds were fixed for testing. For $\mathrm{IoU}@0.50$ localization, precision and recall were reported at Ultralytics' split-specific precision--recall operating point, with an NMS IoU threshold of 0.70.

For each classical and statistical method, detector geometry and structural settings were selected using validation data, while test localization precision and recall were reported at each method's $F_1$-optimal operating point, consistent with the YOLO evaluation.

Hyperparameter configurations for Hough, Radon, and Gaussian PSF are available in the public GitHub repository.
\begin{table*}[t]
\centering
\scriptsize

% ============================================================
% TABLE 2: IMAGE-LEVEL CLASSIFICATION
% ============================================================
\begin{minipage}[t]{0.485\textwidth}
\centering
\setlength{\tabcolsep}{2.2pt}

\refstepcounter{table}
\label{tab:image-level-domain-shift}
{\footnotesize
\textsc{TABLE \thetable}\\
Image-level classification performance for within-source and
leave-one-source-out evaluation.
$\Delta F_1=F_{1,\mathrm{LOO}}-F_{1,\mathrm{within}}$.
}

\vspace{1.5mm}

\resizebox{\linewidth}{!}{%
\begin{tabular}{@{}llrrr@{\hspace{4pt}}rrr@{\hspace{4pt}}r@{}}
\toprule
\multirow{2}{*}{Dataset} &
\multirow{2}{*}{Method} &
\multicolumn{3}{c}{\shortstack{Within-source\\(in-domain)}} &
\multicolumn{3}{c}{\shortstack{Leave-one-out\\(out-of-domain)}} &
\multirow{2}{*}{$\Delta F_1$} \\
\cmidrule(lr){3-5}\cmidrule(lr){6-8}
& & P & R & $F_1$ & P & R & $F_1$ & \\
\midrule

\multirow{4}{*}{StreakMind}
& YOLO26L
& 1.000 & 0.976 & \textbf{0.988}
& 0.923 & 0.857 & \textbf{0.889}
& $-0.099$ \\
& Hough
& 0.756 & 0.738 & 0.747
& 0.447 & 1.000 & 0.618
& $-0.129$ \\
& Radon
& 0.607 & 0.810 & 0.694
& 0.577 & 0.976 & 0.726
& $+0.032$ \\
& Gaussian PSF
& 0.589 & 0.786 & 0.673
& 0.506 & 1.000 & 0.672
& $-0.001$ \\

\midrule

\multirow{4}{*}{UAEMMN}
& YOLO26L
& 0.961 & 1.000 & \textbf{0.980}
& 0.800 & 0.919 & \textbf{0.855}
& $-0.125$ \\
& Hough
& 0.519 & 0.946 & 0.670
& 0.504 & 0.851 & 0.633
& $-0.037$ \\
& Radon
& 0.563 & 0.959 & 0.710
& 0.507 & 0.919 & 0.654
& $-0.056$ \\
& Gaussian PSF
& 0.704 & 0.770 & 0.735
& 0.634 & 0.351 & 0.452
& $-0.283$ \\

\midrule

\multirow{4}{*}{StreaksYoloDataset}
& YOLO26L
& 0.995 & 0.969 & \textbf{0.982}
& 0.946 & 0.622 & 0.751
& $-0.231$ \\
& Hough
& 0.751 & 0.938 & 0.834
& 0.676 & 1.000 & 0.806
& $-0.028$ \\
& Radon
& 0.938 & 0.933 & 0.935
& 0.897 & 0.933 & \textbf{0.915}
& $-0.020$ \\
& Gaussian PSF
& 0.675 & 0.978 & 0.799
& 0.872 & 0.516 & 0.648
& $-0.151$ \\

\midrule

\multirow{4}{*}{Roboflow}
& YOLO26L
& 1.000 & 0.975 & \textbf{0.987}
& 1.000 & 1.000 & \textbf{1.000}
& $+0.013$ \\
& Hough
& 1.000 & 0.900 & 0.947
& 1.000 & 0.875 & 0.933
& $-0.014$ \\
& Radon
& 1.000 & 0.825 & 0.904
& 1.000 & 1.000 & \textbf{1.000}
& $+0.096$ \\
& Gaussian PSF
& 1.000 & 0.900 & 0.947
& 1.000 & 1.000 & \textbf{1.000}
& $+0.053$ \\

\midrule

\multirow{4}{*}{\shortstack[l]{Hubble Asteroid\\Hunter}}
& YOLO26L
& 1.000 & 1.000 & \textbf{1.000}
& 1.000 & 0.900 & 0.947
& $-0.053$ \\
& Hough
& 1.000 & 0.976 & 0.988
& 1.000 & 1.000 & \textbf{1.000}
& $+0.012$ \\
& Radon
& 1.000 & 0.865 & 0.927
& 1.000 & 0.794 & 0.885
& $-0.042$ \\
& Gaussian PSF
& 1.000 & 0.953 & 0.976
& 1.000 & 1.000 & \textbf{1.000}
& $+0.024$ \\

\bottomrule
\end{tabular}%
}

\vspace{-1mm}

\end{minipage}%
\hfill%
% ============================================================
% TABLE 3: IoU@0.50 LOCALIZATION
% ============================================================
\begin{minipage}[t]{0.485\textwidth}
\centering
\setlength{\tabcolsep}{2.2pt}

\refstepcounter{table}
\label{tab:iou50-domain-shift}
{\footnotesize
\textsc{TABLE \thetable}\\
$\mathrm{IoU}@0.50$ precision, recall, and $F_1$ for within-source and
leave-one-source-out evaluation.
$\Delta F_1=F_{1,\mathrm{LOO}}-F_{1,\mathrm{within}}$.
}

\vspace{1.5mm}

\resizebox{\linewidth}{!}{%
\begin{tabular}{@{}llrrr@{\hspace{4pt}}rrr@{\hspace{4pt}}r@{}}
\toprule
\multirow{2}{*}{Dataset} &
\multirow{2}{*}{Method} &
\multicolumn{3}{c}{\shortstack{Within-source\\(in-domain)}} &
\multicolumn{3}{c}{\shortstack{Leave-one-out\\(out-of-domain)}} &
\multirow{2}{*}{$\Delta F_1$} \\
\cmidrule(lr){3-5}\cmidrule(lr){6-8}
& & P & R & $F_1$ & P & R & $F_1$ & \\
\midrule

\multirow{4}{*}{StreakMind}
& YOLO26L
& 0.809 & 0.744 & \textbf{0.775}
& 0.687 & 0.665 & \textbf{0.676}
& $-0.099$ \\
& Hough
& 0.483 & 0.326 & 0.389
& 1.000 & 0.093 & 0.170
& $-0.219$ \\
& Radon
& 0.000 & 0.000 & 0.000
& 0.000 & 0.000 & 0.000
& $0.000$ \\
& Gaussian PSF
& 0.250 & 0.047 & 0.078
& 0.333 & 0.023 & 0.043
& $-0.035$ \\

\midrule

\multirow{4}{*}{UAEMMN}
& YOLO26L
& 0.746 & 0.713 & \textbf{0.729}
& 0.721 & 0.608 & \textbf{0.660}
& $-0.069$ \\
& Hough
& 0.014 & 0.122 & 0.026
& 0.063 & 0.108 & 0.080
& $+0.054$ \\
& Radon
& 0.051 & 0.027 & 0.035
& 0.063 & 0.014 & 0.022
& $-0.013$ \\
& Gaussian PSF
& 0.135 & 0.176 & 0.153
& 0.084 & 0.122 & 0.099
& $-0.054$ \\

\midrule

\multirow{4}{*}{StreaksYoloDataset}
& YOLO26L
& 0.858 & 0.698 & \textbf{0.770}
& 0.675 & 0.418 & \textbf{0.516}
& $-0.254$ \\
& Hough
& 0.570 & 0.418 & 0.482
& 0.013 & 0.013 & 0.013
& $-0.469$ \\
& Radon
& 0.230 & 0.151 & 0.182
& 0.240 & 0.156 & 0.189
& $+0.007$ \\
& Gaussian PSF
& 0.170 & 0.178 & 0.174
& 0.245 & 0.120 & 0.161
& $-0.013$ \\

\midrule

\multirow{4}{*}{Roboflow}
& YOLO26L
& 0.861 & 0.842 & \textbf{0.851}
& 0.483 & 0.184 & \textbf{0.267}
& $-0.584$ \\
& Hough
& 0.172 & 0.355 & 0.232
& 0.391 & 0.118 & 0.182
& $-0.050$ \\
& Radon
& 0.040 & 0.013 & 0.020
& 0.056 & 0.013 & 0.021
& $+0.001$ \\
& Gaussian PSF
& 0.014 & 0.013 & 0.014
& 0.143 & 0.013 & 0.024
& $+0.010$ \\

\midrule

\multirow{4}{*}{\shortstack[l]{Hubble Asteroid\\Hunter}}
& YOLO26L
& 0.708 & 0.676 & \textbf{0.692}
& 0.361 & 0.406 & \textbf{0.382}
& $-0.310$ \\
& Hough
& 0.027 & 0.024 & 0.025
& 0.000 & 0.000 & 0.000
& $-0.025$ \\
& Radon
& 0.053 & 0.071 & 0.061
& 0.084 & 0.071 & 0.077
& $+0.016$ \\
& Gaussian PSF
& 0.006 & 0.006 & 0.006
& 0.019 & 0.018 & 0.018
& $+0.012$ \\

\bottomrule
\end{tabular}%
}

\end{minipage}
\end{table*}

\section{Experimental Results}

\iffull
 \yeye{Roboflow and Hubble $\rightarrow$ label them as ``easy'' somehow on table}
\fi

For \textbf{RQ1}, YOLO26L was the most suitable method for single-source streak detection. It achieved the highest within-source image-level and localization $F_1$ on all five datasets, with means of 0.987 and 0.763, respectively, and localization scores ranging from 0.692 on Hubble to 0.851 on Roboflow.

The non-learned methods were considerably weaker at localization: their mean $F_1$ values were 0.231 for Hough, 0.060 for Radon, and 0.085 for Gaussian PSF. For example, Hough achieved image-level $F_1$ values of 0.947 on Roboflow and 0.988 on Hubble but localization scores of only 0.232 and 0.025. All three non-learned methods achieved similarly high image-level scores on these datasets, but the all-positive test splits did not measure their ability to reject negative images. For practical deployment with large volumes of images, precision and recall should both be around 0.95. Only YOLO26L achieved both image-level precision and
recall above 0.95 across all five within-source tests.

%\yeye{Should stress here, that for practical deployment that operates on large volume of images, high precision adn recall are required, e.g., over 0.95 for both, which makes YOLO26L the only viable option for the ``within source'' setting.}

For \textbf{RQ2}, the evaluated methods generalized inconsistently to unseen sources and generally performed worse than in within-source evaluation. Image-level $F_1$ decreased in 14 of 20 method--source pairs, with a mean $\Delta F_1$ of $-0.052$ (Table~\ref{tab:image-level-domain-shift}). Localization performance was similarly unstable: $\mathrm{IoU}@0.50$ $F_1$ decreased in 13 of 20 pairs, with a mean $\Delta F_1$ of $-0.105$ (Table~\ref{tab:iou50-domain-shift}). In medium and hard
datasets, $F_1$ dropped in 11 of 12 image-level method–source pairs and localization $F_1$ decreased in 9 of 12 pairs and was unchanged in 1.  %\yeye{Can quote these numbers for Hard/Medium dataset only, and explain that easy is too easy with no negative images.}

YOLO26L showed the strongest overall performance but still experienced substantial domain shift. Its image-level $F_1$ decreased on four of five unseen sources. Roboflow was the only exception: its image-level $F_1$ increased from 0.987 to 1.000 ($\Delta F_1=+0.013$), although this result should be interpreted cautiously, as the dataset had an all-positive test split. More importantly, YOLO26L's localization $F_1$ decreased on all five unseen sources by an average of 0.263. The localization losses ranged from $-0.069$ on UAEMMN to $-0.584$ on Roboflow. 

The classical and statistical methods start from a low quality level even for within-source, so they occasionally improved in cross-domain generalization, but these gains were small and inconsistent. 

Overall, none of the evaluated methods demonstrated consistently robust generalization across sources. Although YOLO26L has strong results and is the only method meeting the proposed 0.95 threshold, its results only hold for the \emph{within-source} setting, which \emph{drop substantially for the out-of-domain, leave-one-out setting}, suggesting that it is still infeasible to deploy YOLO26L in a generalized manner to many different observatories. We believe the gap between the within-source and leave-one-out results highlights a genuine practical need for NEO detection and planetary defense, which points to an important direction for future research.  % \yeye{ -- the best method, YOLO26L has only XXX F1 for classification and YYY F1 for locationalization, substantially below the 0.95 level required for practical deployment, while classical and statistical methods scoring even lower in the cross-domain generalization setting, showing a strong opportunity for additional research.}

\iffull Meanwhile, exceptionally poor results from classical methods on StreakMind is caused by... (add info on streakmind dataset)
\fi

\section{Conclusion}

We propose a multi-source astronomical streak benchmark, \bench, that combines 8,376 images from five streak sources to evaluate the generalizability of streak-detection algorithms.

Using \bench, we show that reliable generalization of streak detection across
diverse imaging sources remains an open challenge. We hope the benchmark will
serve as a springboard for further research in this important area of
planetary defense.

\bibliographystyle{IEEEtran}
\bibliography{NEOBench}

\end{document}